\documentclass[10pt, a4paper, twocolumn, twoside]{article} % 10pt font size (11 and 12 also possible), A4 paper (letterpaper for US letter) and two column layout (remove for one column)
\usepackage[english]{babel} % English language hyphenation

\usepackage{microtype} % Better typography

\usepackage{amsmath,amsfonts,amsthm} % Math packages for equations

\usepackage[svgnames]{xcolor} % Enabling colors by their 'svgnames'

\usepackage[small, labelfont=bf, up]{caption} % Custom captions under/above tables and figures

\usepackage{booktabs} % Horizontal rules in tables

\usepackage{lastpage} % Used to determine the number of pages in the document (for "Page X of Total")

\usepackage{graphicx} % Required for adding images

\usepackage{enumitem} % Required for customising lists
\setlist{noitemsep} % Remove spacing between bullet/numbered list elements

\usepackage{sectsty} % Enables custom section titles
\allsectionsfont{\usefont{OT1}{phv}{b}{n}} % Change the font of all section commands (Helvetica)

\usepackage{geometry} % Required for adjusting page dimensions

\usepackage[T1]{fontenc} % Output font encoding for international characters
\usepackage[utf8]{inputenc} % Required for inputting international characters

\usepackage{XCharter} % Use the XCharter font

\usepackage{hyperref}

\usepackage{journals}

\usepackage{fancyhdr} % Needed to define custom headers/footers
\newcommand{\shorttitle}[1]{\fancyhead[CE]{\textsl{#1}}}
\newcommand{\shortauthors}[1]{\fancyhead[CO]{\textsl{#1}}}

\fancypagestyle{firstpage}{ % Page style for the first page with the title
	\fancyhf{}
        \lhead{\vspace*{0.0em} \textit{\footnotesize 21$^{\rm st}$ European Workshop on White
            Dwarfs \\ Castanheira, Campos, Montgomery, eds. \\[-0.2em]
          July 23--27, 2018, Austin, Texas, USA}}
}

\date{}

\newcommand{\authorstyle}[1]{{\large\usefont{OT1}{phv}{b}{n}\color{DarkRed}#1}} % Authors style (Helvetica)

\newcommand{\institution}[1]{{\footnotesize\usefont{OT1}{phv}{m}{sl}\color{Black}#1}} % Institutions style (Helvetica)
\usepackage{titling} % Allows custom title configuration

\newcommand{\HorRule}{\color{DarkGoldenrod}\rule{\linewidth}{1pt}} % Defines the gold horizontal rule around the title

\pretitle{
	\vspace{-30pt} % Move the entire title section up
	\HorRule\vspace{10pt} % Horizontal rule before the title
	\fontsize{16}{12}\usefont{OT1}{phv}{b}{n}\selectfont % Helvetica
	\color{DarkRed} % Text colour for the title and author(s)
}

\posttitle{\par\vskip 10pt} % Whitespace under the title

\preauthor{} % Anything that will appear before \author is printed

\postauthor{ % Anything that will appear after \author is printed
	\vspace{0pt} % Space before the rule
	{\par\HorRule} % Horizontal rule after the title
	\vspace{-10pt} % Space after the title section
}

\newcommand{\newabstract}[1]{
    {\section*{Abstract}
    \bfseries #1}
  }

\usepackage[authoryear]{natbib}
\title{Luminosity and cooling suppression in magnetized white dwarfs} % The article title

\shorttitle{Magnetized white dwarf cooling} % The short article title for page headings

\shortauthors{Bhattacharya, Mukhopadhyay \& Mukerjee} % The short author list for page headings

\author{
        \authorstyle{M.~Bhattacharya,$^1$ B.~Mukhopadhyay,$^2$ and S.~Mukerjee$^2$}
	\newline\newline % Space before institutions
	$^1$\institution{Department of Physics, University of Texas at Austin, Austin, TX 78712, USA; 
          mukul.b@utexas.edu}\\ % Institution 1
	$^2$\institution{Department of Physics, Indian Institute of Science, Bangalore 560012, India; bm@iisc.ac.in}\\ % Institution 2
      }

\begin{document}

\maketitle % Print the title

\thispagestyle{firstpage} % Apply the page style for the first page

%----------------------------------------------------------------------------------------
%	ABSTRACT
%----------------------------------------------------------------------------------------

\newabstract{
We investigate the luminosity and cooling of highly magnetized white dwarfs where cooling occurs by the diffusion of photons. We solve the magnetostatic equilibrium and photon diffusion equations to obtain the temperature and density profiles in the surface layers of these white dwarfs. With increase in field strength, the degenerate core shrinks in volume with a simultaneous increase in the core temperature. For a given white dwarf age and for a fixed interface radius or temperature, the luminosity decreases significantly from $\sim 10^{-6}\, L_{\odot}$ to $10^{-9}\, L_{\odot}$ as the field strength increases from $\sim 10^9$ to $10^{12}\,$G in the surface layers. This is remarkable as it argues that magnetized white dwarfs can remain practically hidden in an observed H--R diagram. We also find that the cooling rates for these highly magnetized white dwarfs are suppressed significantly.
}

%----------------------------------------------------------------------------------------
%	ARTICLE BODY
%----------------------------------------------------------------------------------------

\section{Introduction}
More than a dozen overluminous Type~Ia supernovae have been observed since 2006 (see e.g. \citealt{Howell, Scalzo}), whose significantly high luminosities can be explained by invoking highly super-Chandrasekhar progenitors.
The enormous efficiency of a magnetic field, irrespective of its nature of origin, can explain the existence of significantly super-Chandrasekhar white dwarfs (see e.g. \citealt{mukhoall}, for the current state of this research).
Observations \citep{ferra} indeed confirm that highly magnetized white dwarfs ($B$ $> 10^{6}\,$G) are more massive than non-magnetized white dwarfs. The impact of high magnetic fields not only lies in increasing the limiting mass of white dwarfs but it is also expected to change other properties including luminosity, temperature, cooling rate, etc \citep{mb15}. 

\citet{Mestel} first investigated the cooling of white dwarfs in 1950s in order to estimate the ages of observed white dwarfs. Subsequently, \citet{MestelRud} explored the cooling of white dwarfs and found them to be radiating at the expense of their thermal energy. While the physics of cool white dwarfs was reviewed by \citet{Hansen}, the limitations of Mestel's original theory and its underlying approximations for white dwarf cosmochronology were mentioned later by \citet{font}. The magnetic field effects in white dwarfs become important once the field strength exceeds (see e.g. \citealt{Adam}) the critical field $B_{\rm c}=4.414\times 10^{13}\,$G. Here we estimate the luminosities of magnetized white dwarfs and calculate their corresponding cooling rate by including the contribution of field to the pressure, density, opacity and equation of state (EoS) of white dwarfs. 

\begin{figure}[t]
  \centerline{\includegraphics[angle=0,width=1.00\columnwidth]{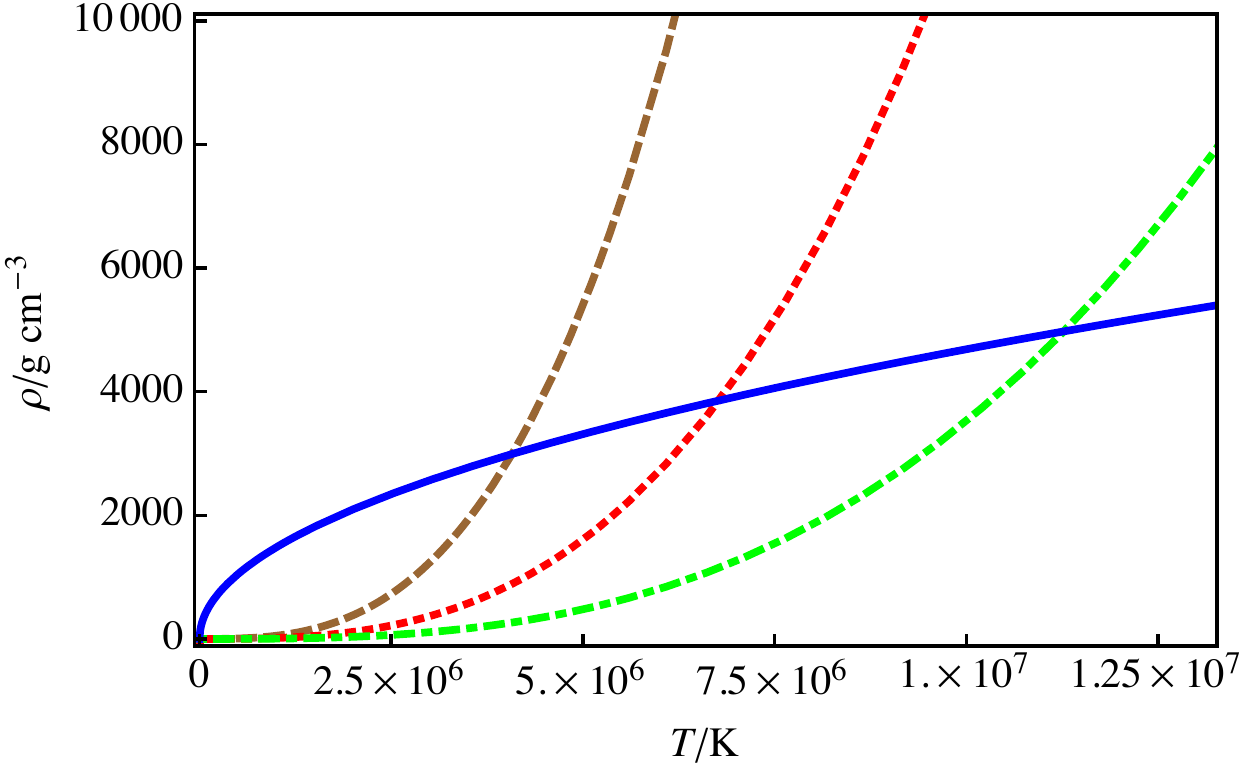}}
  \vspace*{1em}
  \centerline{\includegraphics[angle=0,width=0.95\columnwidth]{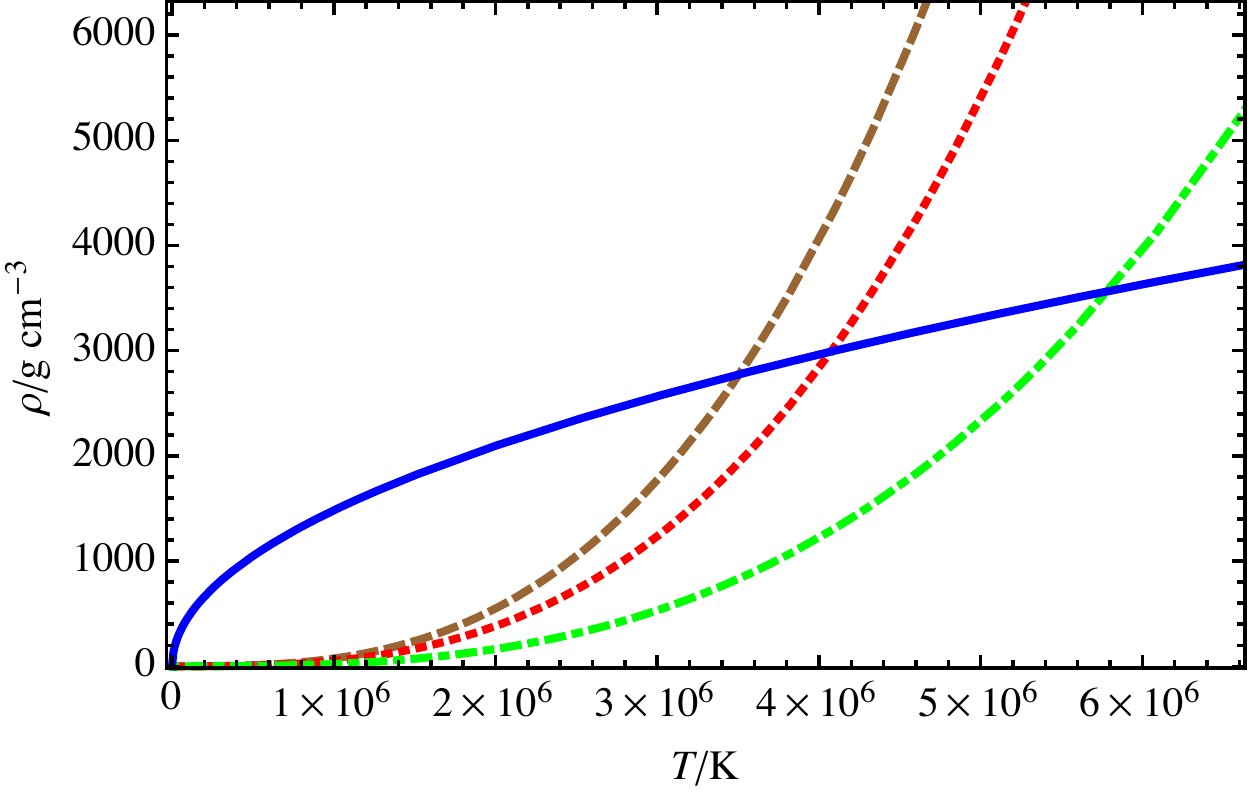}}
  \caption{
  {\it Top panel:} variation of density with temperature for $B\equiv (B_{\rm{s}},B_0) = (10^{12}\,\rm{ G},10^{14}\,\rm{G})$ and different $L$: $10^{-5}\, L_{\odot}$ (dashed line), $10^{-4}\, L_{\odot}$ (dotted line) and $10^{-3}\, L_{\odot}$ (dot-dashed line).  $\rho_{*}$ and $T_{*}$ are obtained from the intersection of the $\rho-T$ profiles with equation (\ref{eqn_rhoT_B}) (solid line).
 {\it Bottom panel:} variation of density with temperature for $L = 10^{-5}\, L_{\odot}$ and different $B$: $(10^{12}\,\rm{G}, 5\times10^{13}\,\rm{G})$ (dashed line), $(10^{12}\,\rm{G}, 10^{14}\,\rm{G})$ (dotted line) and $(10^{12}\,\rm{G}, 5\times10^{14}\,\rm{G})$ (dot-dashed line). The $\rho_{*}$ and $T_{*}$ are obtained from the intersection of the $\rho-T$ profiles with equation (\ref{eqn_rhoT_B}) (solid line).
    }
  \label{fig2}
\end{figure}

\begin{figure}[t]
  \centerline{\includegraphics[angle=0,width=1.00\columnwidth]{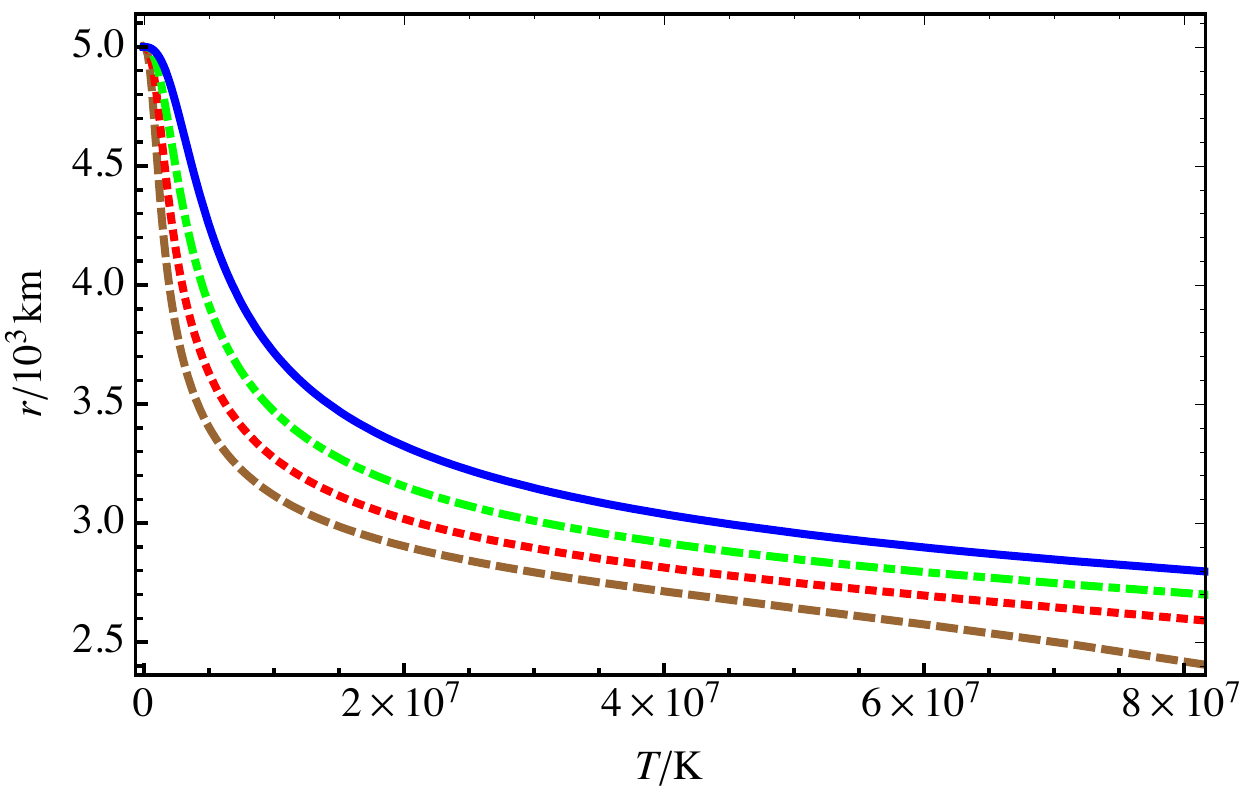}}
  \vspace*{1em}
  \centerline{\includegraphics[angle=0,width=1.00\columnwidth]{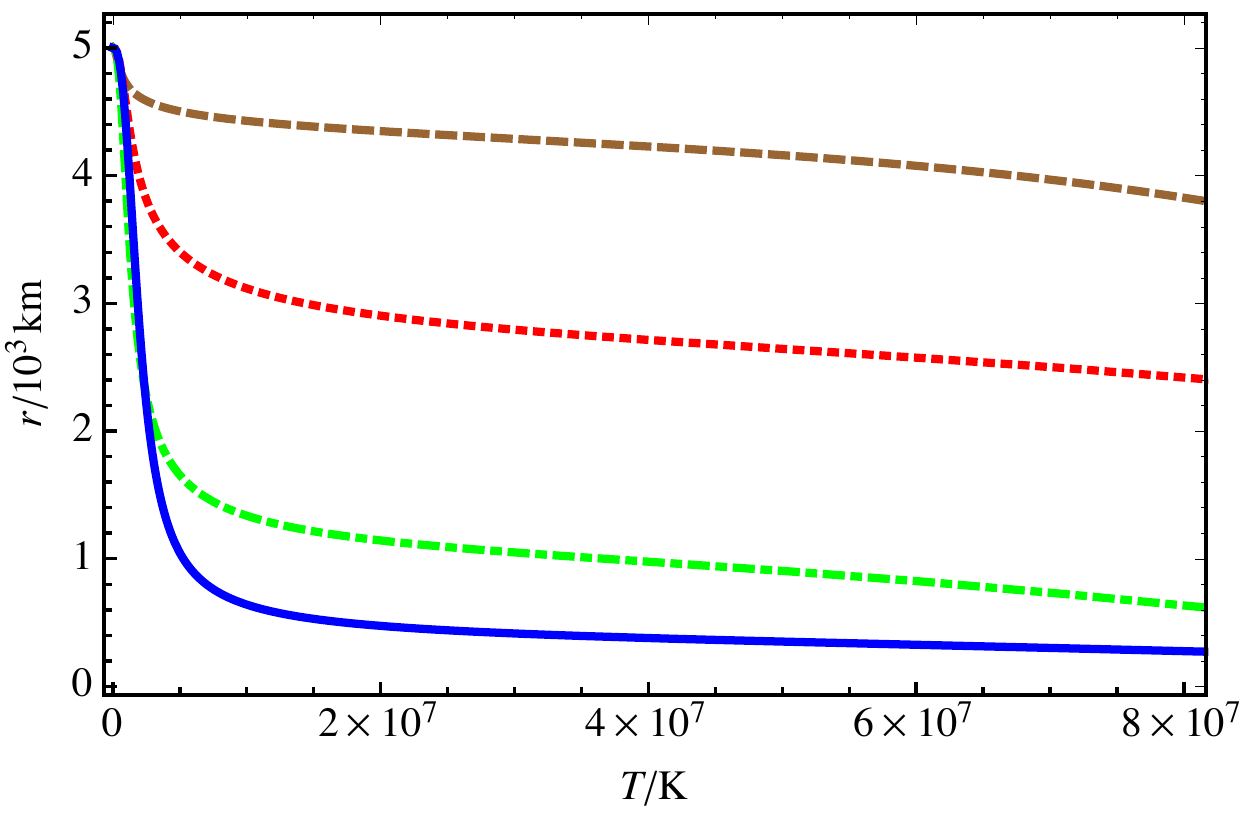}}
  \caption{
  {\it Top panel:} variation of radius with temperature for $B = (10^{12}\,\rm{G}, 10^{14}\,\rm{G})$ and different luminosities: $10^{-5}\, L_{\odot}$ (dashed line), $10^{-4}\, L_{\odot}$ (dotted line), $10^{-3}\, L_{\odot}$ (dot-dashed line) and $10^{-2}\, L_{\odot}$ (solid line).
 {\it Bottom panel:} variation of radius with temperature for $L = 10^{-5}\, L_{\odot}$ and different magnetic fields: $ (10^{11}\, \rm{G}, 10^{14}\, \rm{G})$ (dashed line), $ (10^{12}\, \rm{G}, 10^{14}\, \rm{G})$ (dotted line), $ (10^{12}\, \rm{G}, 5\times10^{14}\, \rm{G})$ (dot-dashed line) and $ (5\times 10^{12}\,\rm{G}, 5\times10^{14}\, \rm{G})$ (solid line).
    }
  \label{fig4}
\end{figure}

\section{Magnetized white dwarf properties}
\label{sec2}
We solve the magnetostatic equilibrium and photon diffusion equations in the presence of a magnetic field ($\vec{B}$) to investigate the temperature profile inside a white dwarf. We perform our calculations for realistic radially varying magnetic fields. The field inside a white dwarf gives rise to magnetic pressure, $P_{B} = {B^2}/{8\pi}$, where $B=\sqrt{\vec{B}.\vec{B}}$, which contributes to the matter pressure and gives rise to the total pressure (see, e.g., \citealt{Sinha}). Moreover, the density has a contribution from the field that is given by $\rho_{B} = {B^2}/{8\pi c^2}$ \citep{Sinha}. The opacity and EoS of the matter are also modified by $\vec{B}$. The magnetostatic equilibrium and photon diffusion equations are
\begin{equation}
\frac{d}{dr}(P+P_{B}) = -\frac{GM}{r^2}(\rho+\rho_{B}),
\label{eqn1}
\end{equation}
and
\begin{equation}
\frac{dT}{dr} = -\frac{3}{4ac}\frac{\kappa(\rho+\rho_{B})}{T^{3}}\frac{L}{4\pi r^2},
\label{eqn2}
\end{equation}
respectively, neglecting magnetic tension terms. In these equations, $P$ is the matter pressure which is same as the core electron degeneracy pressure, $\rho$ is the matter density, $\kappa$ is the radiative opacity, $T$ is the temperature, $a$ is the radiation constant, $c$ is the speed of light in vacuum, $G$ is Newton's gravitational constant, $m(r) \approx M$ is the mass enclosed within radius $r$ in the envelope, and $L$ is the luminosity. 

The opacity due to the bound-free and free-free transitions of electrons \citep{Shapiro} for a non-magnetized white dwarf is approximated with Kramers' formula, $\kappa = \kappa_{0}\rho T^{-3.5}$, where $\kappa_{0} = 4.34\times10^{24} Z (1+X)\,$$\rm{ cm^{2}g^{-1}}$ and $X$ and $Z$ are the mass fractions of hydrogen and heavy elements (elements other than hydrogen and helium) in the stellar interior, respectively. For a typical white dwarf, $X=0$, and we assume that the mass fraction of helium $Y=0.9$ and $Z=0.1$ for simplicity. For the large fields considered here, the radiative opacity variation with $B$ can be modelled similarly to neutron stars as $\kappa = \kappa_{B} \approx 5.5\times10^{31} \rho T^{-1.5}B^{-2}\,$$\rm{cm^{2}g^{-1}}$ \citep{PotYak,VenPot}. We use a profile proposed by \citet{Bandyopadhyay} to model $B$ as a function of $\rho$ and capture the variation of field strength irrespective of the other complicated effects (including the field geometry) that might be involved, 
\begin{equation}
B\left(\frac{\rho}{\rho_{0}}\right) = B_{\rm{s}} + B_{0}\left[1-\rm{exp}{\left(-\eta \left(\frac{\rho}{\rho_{0}}\right)^{\gamma}\right)}\right],
\label{eqn5}
\end{equation} 
where $B_{\rm{s}}$ is the surface magnetic field, $B_{0}$ (similar to the central field) is a parameter with the dimension of $B$. $\eta$ and $\gamma$ are parameters that determine how the field strength reduces from the core to the surface. We choose $\rho_{0} \approx 0.1\rho_{\rm{c}}$, where $\rho_{\rm{c}}$ is the central density, and set $\eta = 0.8$, $\gamma = 0.9$ and $\rho_{0} = 10^{9}\, \rm{g\,cm^{-3}}$ for all our calculations. We neglect complicated effects such as offset dipoles and magnetic spots that can arise from more complex field structures (see e.g. \citealt{Max,Ven}).

\begin{table}[t]
\begin{center}
%\small
\caption{ Variation of luminosity with magnetic field for fixed $r_{*}=0.9978\,R$}
\begin{tabular}{cccccccccccccccccccccc}
\hline
\hline
\centering
$B/\rm{G}=(B_{\rm{s}}/G,B_0/G)\,$ & $L/L_{\odot}$ &  $T_{*}/\rm{K}$ \\ \hline
$(0, 0)$                 & $1.00\times10^{-5}$ & $2.332\times10^{6}$ \\ \hline
$(10^9,6\times10^{13})$ & $2.53\times10^{-7}$ & $4.901\times10^{5}$ \\ \hline
%$(2\times10^{9},4\times10^{13})$ & $2.07\times10^{-8}$ & $2.737\times10^{5}$ \\ \hline
$(5\times10^{9},2\times10^{13})$ & $3.96\times10^{-8}$ & $3.262\times10^{5}$ \\ \hline
$(10^{10},10^{13})$ & $1.02\times10^{-6}$ & $7.189\times10^{5}$ \\ \hline
%$(2\times10^{10},6\times10^{12})$ & $1.22\times10^{-6}$ & $7.616\times10^{5}$ \\ \hline
$(2\times10^{10},8\times10^{12})$ & $4.40\times10^{-9}$ & $2.063\times10^{5}$ \\ \hline
$(5\times10^{10},4\times10^{12})$ & $2.59\times10^{-8}$ & $3.185\times10^{5}$ \\ \hline
%$(10^{11},2\times10^{12})$ & $1.09\times10^{-6}$ & $7.721\times10^{5}$ \\ \hline
$(5\times10^{11},10^{12})$ & $2.93\times10^{-9}$ & $2.206\times10^{5}$ \\ \hline
\hline
\label{table5}
\end{tabular}
\end{center}
\end{table}

\begin{table}[t]
\begin{center}
%\small
\caption{ Variation of luminosity with magnetic field for fixed $T_{*}=2.332\times10^{6}\,\rm{K}$}
\begin{tabular}{cccccccccccccccccccccc}
\hline
\hline
\centering
$B/\rm{G}=(B_{\rm{s}}/G,B_0/G)\,$ & $L/L_{\odot}$ & $r_{*}/R$ \\ \hline
$(0, 0)$ & $1.00\times10^{-5}$ & $0.9978$ \\ \hline
$(10^{11}, 5\times10^{14})$ & $1.26\times10^{-6}$ & $0.6910$ & \\ \hline
%$(2\times10^{11}, 5\times10^{14})$ & $6.77\times10^{-7}$ & $0.5830$ \\ \hline
$(5\times10^{11}, 5\times10^{14})$ & $2.98\times10^{-7}$ & $0.4342$ \\ \hline
%$(10^{12}, 10^{14})$ & $7.93\times10^{-7}$ & $0.7131$ \\ \hline
$(10^{12}, 5\times10^{14})$ & $1.60\times10^{-7}$ & $0.3326$ \\ \hline
$(2\times10^{12}, 10^{14})$ & $4.26\times10^{-7}$ & $0.6236$ \\ \hline
%$(2\times10^{12}, 5\times10^{14})$ & $8.57\times10^{-8}$ & $0.2491$ \\ \hline
$(5\times10^{12}, 10^{14})$ & $1.87\times10^{-7}$ & $0.5055$ \\ \hline
$(5\times10^{12}, 5\times10^{14})$ & $3.76\times10^{-8}$ & $0.1698$ \\ \hline
\hline
\label{table6}
\end{tabular}
\end{center}
\end{table}

Dividing equations (\ref{eqn1}) and (\ref{eqn2}), we obtain
\begin{equation}
\frac{d}{dT}\left(P+P_B\right) = \frac{4ac}{3} \frac{4\pi GM}{L} \frac{T^{3}}{\kappa}.
\label{eqn3}
\end{equation}

The contribution of $\rho_B$ to the matter density cannot be ignored for $B_{\rm{s}} > 10^{12}\,$G (see \citealt{heanselbook} for details), and the EoS for the degenerate core after including the quantum mechanical effects depends on the field strength \citep{VenPot}. Equating the electron pressure for the non-relativistic electrons on both sides of the interface gives 
\begin{equation}
\rho_{*}(B_{*}) \approx (1.482\times10^{-12})\, T_{*}^{1/2}B_{\rm s}
%{\rm{\,g\,cm^{-3}\,K^{-1/2}\,G^{-1}}}
\label{eqn_rhoT_B}
\end{equation}
\citet{Tremblay} showed that unlike neutron stars, changes in transverse conduction rates in white dwarfs due to magnetic fields do not affect the cooling process as thermal conduction takes place only in the stellar interior and the insulating region is non-degenerate. In this work, we consider white dwarfs with isothermal core and mass $M=M_{\odot}$ corresponding to radius $R=5000\,$km from Chandrasekhar's relation for white dwarfs (\citealt{Chandra1, Chandra2}). 

We consider a realistic density dependent magnetic field profile such that the field strength decreases from the core to the surface for spherically symmetric white dwarfs. It should be noted that the central and surface fields (and hence corresponding $B_0$ and $B_{\rm{s}}$ in equation \ref{eqn5}) are chosen keeping stability criteria in mind. \citet{brait} earlier argued that the magnetic energy should be well below the gravitational energy in order to form a stable white dwarf. We fix the radius ($R=5000\,$km) throughout even though this need not be the case for all chosen fields. 

As we are interested in the surface layers that are non-degenerate, we can substitute $P$ in terms of $\rho$ in equation (\ref{eqn3}) by the ideal gas EoS to obtain
\begin{eqnarray}
(5.938\times10^{7})\, \rho + (5.938\times10^{7})\, T \frac{d\rho}{dT}\nonumber \\ 
+ 0.0796 B \frac{dB}{d\rho} \frac{d\rho}{dT} =  \frac{(9.218\times10^{-9})}{L} \frac{T^{4.5}}{\rho}B^2.\nonumber\\
% {\rm{cm^{2}\,s^{-2}\,K^{-1}}}
% {\rm{cm^{2}\,s^{-2}\,K^{-1}}}
% {\rm{g^{2}\,cm^{-1}\,s^{-3}\,K^{-5.5}}}
\label{eqn7}
\end{eqnarray}
From equation (\ref{eqn2}), we further have
\begin{equation}
\frac{dr}{dT} = -(6.910\times10^{-35}) \frac{T^{4.5}B^2}{\rho\left(\rho + \frac{B^2}{2.261\times10^{22}}\right)} \frac{r^{2}}{L}.
% {\rm{g^{2}\,cm^{-4}\,s^{-1}\,K^{-5.5}}}
\label{eqn8}
\end{equation}
Equations (\ref{eqn7}) and (\ref{eqn8}) are simultaneously solved with surface boundary conditions: $\rho(T_{\rm{s}}) = 10^{-10}\, \rm{g\,cm^{-3}}$ and $r(T_{\rm{s}}) = R = 5000\,$km. 

Once we obtain the $\rho-T$ and $r-T$ profiles for the given boundary conditions, $T_{*}$ and $\rho_{*}$ can be obtained by solving for the $\rho-T$ profile along with equation (\ref{eqn_rhoT_B}), as shown in Fig. \ref{fig2}. After obtaining $T_{*}$, we can also find $r_{*}$ from the $r-T$ profile. We find that the interface moves inwards with an increase of field strength and an increase of luminosity. As opposed to the non-magnetized white dwarfs, the $r - T$ profile is no longer linear for any $L$ (see Fig. \ref{fig4}). We find that the temperature-fall rate near the surface increases with luminosity and decreases with field strength. The density $\rho_*$ also increases with the increase of $L$ or $B$, as $\rho_* \propto T_*^{1/2}B$ from equation (\ref{eqn_rhoT_B}).

\begin{figure}[t]
  \centerline{\includegraphics[angle=0,width=1.00\columnwidth]{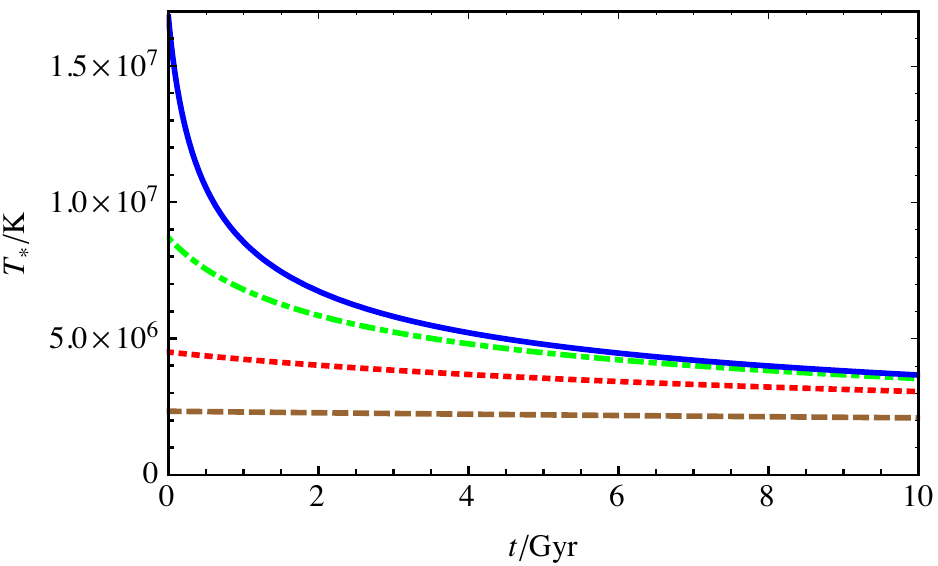}}
  \vspace*{1em}
  \centerline{\includegraphics[angle=0,width=0.95\columnwidth]{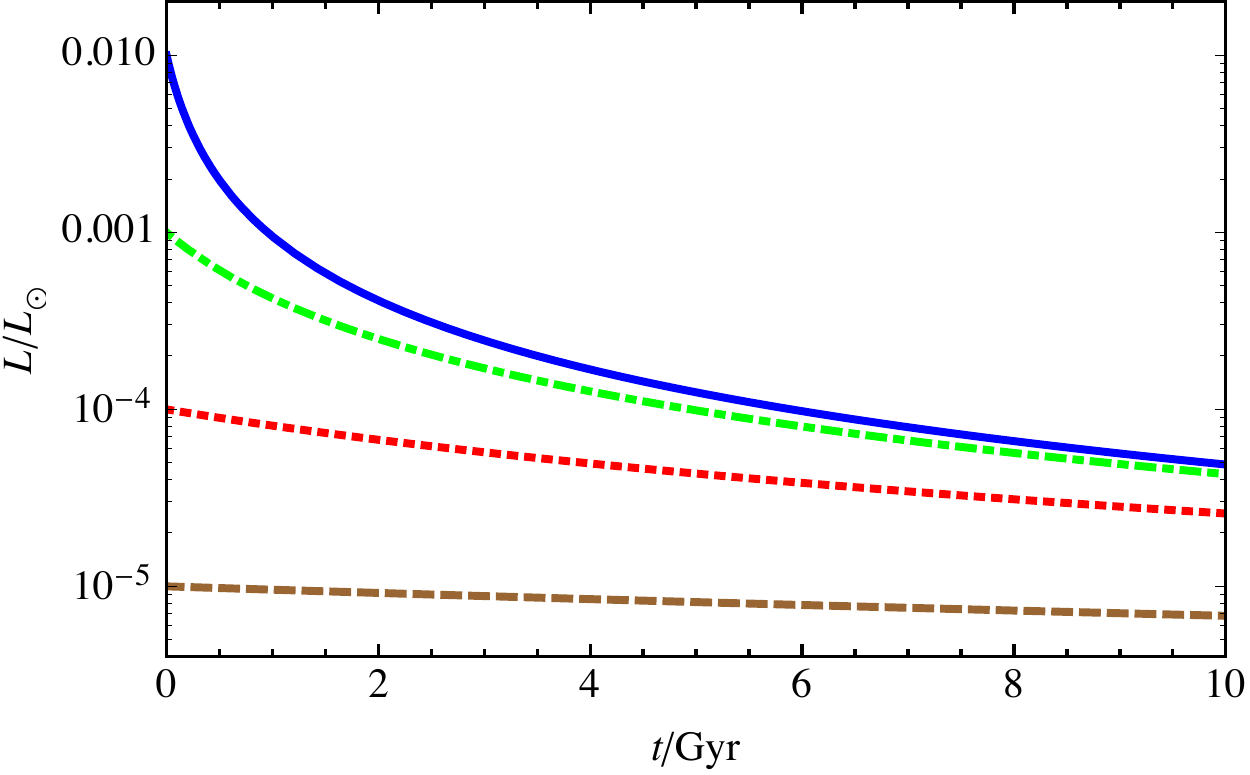}}
  \caption{
  {\it Top panel:} variation of interface temperature with time for a non-magnetized white dwarf with different initial luminosities: $10^{-5} L_{\odot}$ (dashed line), $10^{-4} L_{\odot}$ (dotted line), $10^{-3} L_{\odot}$ (dot-dashed line) and $10^{-2} L_{\odot}$ (solid line).
 {\it Bottom panel:} variation of luminosity with time for a non-magnetized white dwarf with different initial luminosities: $10^{-5} L_{\odot}$ (dashed line), $10^{-4} L_{\odot}$ (dotted line), $10^{-3} L_{\odot}$ (dot-dashed line) and $10^{-2} L_{\odot}$ (solid line).
    }
  \label{fig5}
\end{figure}

\begin{table*}
\begin{center}
%\small
\caption{ Change in $T_{*}$ with time due to the presence of a magnetic field for fixed $r_{*}=0.9978\,R$}%and $T_{s}=(L/4\pi R^2 \sigma)^{1/4}$}
\begin{tabular}{cccccccccccccccccccccc}
\hline
\hline
\centering
$B/\rm{G}=(B_{\rm{s}}/\rm{G},B_0/\rm{G})\,$ & $T_{*,\rm{in}}/\rm{K}\,$ & $L_{\rm{in}}/L_{\odot}$ & $L(T)/{\rm erg\ s^{-1}}$ & $T_{*,\rm{pr}}/\rm{K}$ \\ \hline
$(0, 0)$                 & $2.332\times10^{6}$ & $1.00\times10^{-5}$ & $2.013\times10^{6}T^{3.500}$ & $2.223\times10^{6}$ \\ \hline
$(10^{9},6\times10^{13})$ & $4.901\times10^{5}$ & $2.53\times10^{-7}$ & $2.288\times10^{4}T^{3.971}$ & $4.874\times10^{5}$ \\ \hline
%$(2\times10^{9},4\times10^{13})$ & $2.737\times10^{5}$ & $2.07\times10^{-8}$ & $1.551\times10^{3}T^{4.172}$ & $2.735\times10^{5}$ \\ \hline
$(5\times10^{9},2\times10^{13})$ & $3.262\times10^{5}$ & $3.96\times10^{-8}$ & $1.665\times10^{3}T^{4.160}$ & $3.258\times10^{5}$ \\ \hline
$(10^{10},10^{13})$ & $7.189\times10^{5}$ & $1.02\times10^{-6}$ & $2.951\times10^{4}T^{3.943}$ & $7.081\times10^{5}$ \\ \hline
%$(2\times10^{10},6\times10^{12})$ & $7.616\times10^{5}$ & $1.22\times10^{-6}$ & $2.474\times10^{4}T^{3.952}$ & $7.488\times10^{5}$ \\ \hline
$(2\times10^{10},8\times10^{12})$ & $2.063\times10^{5}$ & $4.40\times10^{-9}$ & $1.627\times10^{2}T^{4.328}$ & $2.062\times10^{5}$ \\ \hline
$(5\times10^{10},4\times10^{12})$ & $3.185\times10^{5}$ & $2.59\times10^{-8}$ & $3.277\times10^{2}T^{4.263}$ & $3.182\times10^{5}$ \\ \hline
%$(10^{11},2\times10^{12})$ & $7.721\times10^{5}$ & $1.09\times10^{-6}$ & $7.099\times10^{3}T^{4.032}$ & $7.606\times10^{5}$ \\ \hline
$(5\times10^{11},10^{12})$ & $2.206\times10^{5}$ & $2.93\times10^{-9}$ & $2.407\times10^{1}T^{4.428}$ & $2.206\times10^{5}$ \\ \hline
\hline
\label{table7}
\end{tabular}
\end{center}
\end{table*}

\begin{table*}
\begin{center}
%\small
\caption{ Change in $T_{*}$ with time due to the presence of a magnetic field for fixed $T_{*}=2.332\times10^{6}\,\rm{K}$} %and $T_{s}=(L/4\pi R^2 \sigma)^{1/4}$}
\begin{tabular}{cccccccccccccccccccccc}
\hline
\hline
\centering
$B/\rm{G}=(B_{\rm{s}}/\rm{G},B_0/\rm{G})\,$ & $T_{*,\rm{in}}/\rm{K}\,$ & $L_{\rm{in}}/L_{\odot}$ & $L(T)/{\rm erg\ s^{-1}}$ & $T_{*,\rm{pr}}/\rm{K}$ \\ \hline
$(0, 0)$                & $2.332\times10^{6}$ & $1.00\times10^{-5}$ & $2.013\times10^{6}T^{3.500}$ & $2.223\times10^{6}$ \\ \hline
$(10^{11},5\times10^{14})$ & $2.332\times10^{6}$ & $1.26\times10^{-6}$ & $5.901*10^{-2}T^{4.541}$ & $2.317\times10^{6}$ \\ \hline
%$(2\times10^{11},5\times10^{14})$ & $2.332\times10^{6}$ & $6.77\times10^{-7}$ & $2.996\times10^{-2}T^{4.545}$ & $2.324\times10^{6}$ \\ \hline
$(5\times10^{11},5\times10^{14})$ & $2.332\times10^{6}$ & $2.98\times10^{-7}$ & $1.317\times10^{-2}T^{4.545}$ & $2.328\times10^{6}$ \\ \hline
%$(10^{12},10^{14})$ & $2.332\times10^{6}$ & $7.93\times10^{-7}$ & $3.715\times10^{-2}T^{4.541}$ & $2.323\times10^{6}$ \\ \hline
$(10^{12},5\times10^{14})$ & $2.332\times10^{6}$ & $1.60\times10^{-7}$ & $7.072\times10^{-3}T^{4.545}$ & $2.330\times10^{6}$ \\ \hline
$(2\times10^{12},10^{14})$ & $2.332\times10^{6}$ & $4.26\times10^{-7}$ & $1.882\times10^{-2}T^{4.545}$ & $2.327\times10^{6}$ \\ \hline
%$(2\times10^{12},5\times10^{14})$ & $2.332\times10^{6}$ & $8.57\times10^{-8}$ & $3.474\times10^{-3}T^{4.552}$ & $2.331\times10^{6}$ \\ \hline
$(5\times10^{12},10^{14})$ & $2.332\times10^{6}$ & $1.87\times10^{-7}$ & $7.583\times10^{-3}T^{4.552}$ & $2.330\times10^{6}$ \\ \hline
$(5\times10^{12},5\times10^{14})$ & $2.332\times10^{6}$ & $3.76\times10^{-8}$ & $1.567*10^{-3}T^{4.550}$ & $2.332\times10^{6}$ \\ \hline
\hline
\label{table8}
\end{tabular}
\end{center}
\end{table*}

\vspace*{-0.5em}

\section{Luminosity variation with field strength}
\label{sec3}
Here we determine how the luminosity of a white dwarf changes as the field strength increases such that
\vspace{0.05 in}
\\(i) the interface radius for a magnetized white dwarf is the same as that for a non-magnetized white dwarf, $r_{*,B\neq0}=r_{*,B=0}$, and
\\(ii) the interface temperature for a magnetized white dwarf is the same as that for a non-magnetized white dwarf, 
$T_{*,B\neq0}=T_{*,B=0}$. 
\vspace{0.1in} 

The motivation for fixing $r_*$ or $T_*$ is to better constrain the individual components (gravitational, thermal and magnetic) of the conserved total energy of the magnetized white dwarf. For fixed $r_{*}$, we assume that the increase in field energy is compensated by an equal decrease in the degenerate core thermal energy while the gravitational potential energy remains unaffected due to fixed $r_{*}$ and $R$. This is also justified by the reduction in $T_{*}$ (and thereby $L$) with increase in $B$ (see Table \ref{table5}). For fixed $T_{*}$, we assume that the increase in field energy is compensated by an equal decrease in gravitational potential energy of the white dwarf whereas the thermal energy is unaffected due to fixed core temperature $T_{\rm core}=T_{*}$.

\subsection{Fixed interface radius}
\label{sec3.1}
We assume a field profile as given by equation (\ref{eqn5}) to find the variation of luminosity with change in $B_{\rm{s}}$ and $B_{0}$ such that the interface radius is same as for the non-magnetic case. For $B=0$ and $L=10^{-5}\, L_{\odot}$, we have $r_{*}=0.9978\, R$, $\rho_{*}=170.7 \,\rm{g\,cm^{-3}}$ and $T_{*} = 2.332\times10^{6}\,$K (see Table 1). Using the same boundary conditions as in section \ref{sec2} we solve equations (\ref{eqn7}) and (\ref{eqn8}) but this time vary $L$ in order to fix $r_{*}=0.9978\,R$. 

Table \ref{table5} shows that $L$ and $T_{*}$ both decrease as the field strength increases. However, the change is appreciable only when $B_{\rm{s}} \ge 10^{10}\,$G or $B_{0} \ge 10^{13}\,$G with $L$ becoming quite low $L \approx 10^{-6}\,L_{\odot}$, and lower for white dwarfs with $(B_{\rm{s}},\,B_{0}) = (2\times10^{10}\,\rm{G},\,7\times10^{12}\,\rm{G})$ and higher. The considerable reduction in $L$ makes it difficult to detect such highly magnetized white dwarfs.

\subsection{Fixed interface temperature}
\label{sec3.2}
Now we solve equations (\ref{eqn7}) and (\ref{eqn8}) as done in section \ref{sec2}, but this time vary $L$ to get $T_{*} = 2.332\times10^{6}\,$K with the same boundary conditions as in section \ref{sec2}. We find that for $T_{*}$ to be unchanged, $L$ has to decrease as $B$ increases. Moreover from Table \ref{table6}, we see that $L$ becomes very small when $B_{\rm{s}} > 2\times10^{11}$ and $B_{0} \ge 2\times10^{14}\,$G. We also see that $r_{*}$ decreases with increase in field strength.

\section{Magnetized white dwarf cooling and temperature profile}
\label{sec4}
Here we briefly discuss how the cooling time-scale of a white dwarf can be evaluated once we know the $L-T$ relation. We first estimate $L-T$ relations for magnetized white dwarfs by fitting power laws of the form $L = \alpha T^{\gamma}$ for different field strengths. Using those $L-T$ relations, we then estimate the cooling over time to find the present interface temperature, $T_{*, \rm{pr}}$, from the initial interface temperature $T_{*, \rm{in}}$ for white dwarf age $\tau = 10\,$Gyr.

\subsection{White dwarf cooling timescale}
\label{sec4.1}
The ion thermal energy and the rate at which it is transported to the surface to be radiated depends on the specific heat, which depends significantly on the physical state of the ions in the core. The white dwarf cooling rate $-dU/dt$ can be equated to $L$ to write (\citealt{Shapiro})
\begin{equation}
L=-\frac{d}{dt} \int c_{\rm{v}} dT = (2\times10^{6}) \frac{Am_{\mu}}{M_{\odot}} T^{7/2},
\label{eqnx}
\end{equation}
where $c_{\rm{v}}$ is the specific heat at constant volume and $A$ is the atomic weight. For $T \gg T_{\rm{g}}$ (where $T_{\rm{g}}$ corresponds to the temperature at which the ion kinetic energy exceeds its vibrational energy), $c_{\rm v} \approx 3k_{\rm{B}}/2$, where $k_{\rm{B}}$ is the Boltzmann constant. This then gives 
\begin{eqnarray}
\left(T^{-5/2}-{T_{0}}^{-5/2}\right) = (3.3\times10^{6}) \frac{Am_{\mu}}{M_{\odot}} \frac{(t-t_{0})}{k_{\rm{B}}} \nonumber \\
= (2.4058\times10^{-34})\tau,
\label{eqn9}
\end{eqnarray}
where $T_{0}$ is the initial temperature, $T$ is the present temperature (at time $t$) and $\tau = t-t_{0}$ is the white dwarf age. 

We first estimate $T$ for $T_{*}=T_{0}$ and $\tau = 10\,$Gyr = $3.1536\times10^{17}\,$s. The top panel of Fig. \ref{fig5} shows that the cooling at the interface is significant only for higher luminosities ($L\ge 10^{-3} L_{\odot}$) and that white dwarfs spend most of the time near their present temperature. The bottom panel of Fig. \ref{fig5} shows that even after $10\,$Gyr, $L$ reduces only by a single order of magnitude, explaining why many white dwarfs have not yet faded away from view, even though their initial luminosities may have been quite low. 

It should be noted that convection might also result in shorter cooling time-scales due to more efficient energy transfer but it has been shown not to be significant \citep{LambHorn,FonHorn} to a first-order approximation. This is due to the fact that convection does not influence the cooling time until the convection zone base reaches the degenerate thermal energy reservoir and couples the surface with the reservoir. This coupling only occurs for surface temperatures much lower than what we have considered here. \citet{Tremblay} showed that the convective energy transfer is significantly hampered when the magnetic pressure dominates over the thermal pressure. 

In the presence of a magnetic field, the state of the ionic core and its specific heat are affected. The relevant parameter to quantify this effect is 
\begin{equation}
b = \frac{\omega_{B}}{\omega_{p}},
\end{equation}
where 
\begin{equation}
\omega_{B} = \frac{ZeB}{Mc}, \,\,\,{\rm and}\,\,\, \omega_{p} = \sqrt{\frac{4\pi Z^{2}e^{2}n}{M}},
\end{equation}
are the ion cyclotron and ion plasma frequencies, respectively. Here $n$ is the number density of the ions,
$e$ is the electric charge and $\omega_p$ is the effective Debye frequency of the ionic lattice.

\citet{Baiko} studied the effect of magnetic fields on Body Centered Cubic (BCC) Coulomb lattice and concluded that there is an appreciable change of the specific heat only for $b \gg 1$ unless $T \ll \theta_D$ (Debye temperature). For almost all the white dwarfs that we consider in this work, $B < 10^{12}\,$G at the interface, that is $b \le 1$. Moreover, the interface temperature is not significantly smaller than $\theta_D$. So, it is justified to work with a specific heat appropriate for a non-magnetized system inspite of the presence of a magnetic field.

\subsection{Fixed interface radius} 
\label{sec4.2}
We first estimate the $L =\alpha T^{\gamma}$ relations for different field strengths. From Table \ref{table5}, we know the initial interface luminosity prior to cooling, $L_{*,\rm{in}}$, and the corresponding initial interface temperature, $T_{*,\rm{in}}$, for different field strengths. Using these for the cooling evolution (equation \ref{eqnx}), we calculate the present interface temperature, $T_{*,\rm{pr}}$, for different $B$ and $r_{*}=0.9978\, R$, as shown in Table \ref{table7}. We find that with the increase in field strength, $L$ gradually decreases as the coefficient $\alpha$ in the $L=\alpha T^{\gamma}$ relation decreases and the exponent $\gamma$ increases. Moreover, increasing $B$ results in slower cooling of the white dwarf.

\subsection{Fixed interface temperature}
\label{sec4.3}
As earlier, the $L=\alpha T^{\gamma}$ relations for different $B$ are estimated and $L_{*,\rm{in}}$ for different fields are obtained from Table \ref{table6}. We then calculate $T_{*,\rm{pr}}$ for the different $B$ and $T_{*} = 2.332\times10^{6}\,$K using equation (\ref{eqnx}), as shown in Table \ref{table8}. An increase in the field strength results in a decrease in the coefficient $\alpha$ and increase in the exponent $\gamma$ in the $L=\alpha T^{\gamma}$ relation, as shown in Table \ref{table8}. Similar to the fixed $r_*$ case, the cooling rate decreases appreciably with an increase in field strength for $B_{\rm{s}} \ge 5\times10^{11}\,$G and $B_{0} \ge 5\times10^{14}\,$G.

%\vspace*{-0.5em}
\section{Summary \& Conclusions}
\label{sec5}
In this paper, we studied the luminosity and cooling of magnetized white dwarfs taking into account the field effects on the EoS, opacity, thermal conductivity and other observables. We have computed the luminosity variation with varying field strength and evaluated the corresponding cooling timescales for white dwarfs with the same interface radius or temperature as their non-magnetic counterparts. We have found that for a given white dwarf age, the luminosity is suppressed with an increase in field strength, in addition to a reduction of the cooling rates. This apparent correlation between luminosity and field strength is found for higher fields only, $(B_{\rm{s}},B_0) > (10^9,10^{13})\,$G. 

Observations indeed suggest that stronger fields with $B_{\rm{s}} < 10^6\,$G correspond to lower $T_{\rm{s}}$ and hence smaller luminosity \citep{ferra}. From the number distribution of white dwarfs with field strength (\citealt{ferra}), we find that there are fewer white dwarfs observed with larger fields. Hence, extrapolating this trend, we expect that our results would be in accordance with the observations when white dwarfs with higher field strength ($B_{\rm{s}} > 10^9\,$G) are observed in future. For a similar gravitational energy, an increase in magnetic energy necessarily requires decrease in thermal energy for white dwarfs to be in equilibrium, resulting in a corresponding decrease in luminosity. 

It should be noted that understanding the evolution and structure of a white dwarf is a complicated time-dependent nonlinear problem. As a result, our findings should be confirmed based on more rigorous computations, without assuming that the core is perfectly isothermal, cooling process will be self-similar up to $10\,$Gyr, etc. Nevertheless, we have found that the luminosity could be as low as about $10^{-8}\,L_\odot$ for a white dwarf with central field $\sim 5\times 10^{14}\,$G and surface field $\sim 5\times10^{12}\,$G, for the same interface temperature as non-magnetized white dwarfs. The luminosity for a fixed interface radius could be even lower, $L \approx 10^{-9}\,L_\odot$, for central and surface field strengths of about $10^{12}\,$G and $5\times10^{11}\,$G, respectively. Therefore, such white dwarfs, while expected to be present in the Universe, would be virtually invisible to us, and perhaps lie in the lower left-hand corner in the H--R diagram.

%----------------------------------------------------------------------------------------
%	BIBLIOGRAPHY
%----------------------------------------------------------------------------------------

% There are two ways to include references. The first uses bibtex and
% is recommended. For this case uncomment the following line. 
\bibliography{papers}

\end{document}